\documentclass[%
mathleft,%
final,%
]{an}
\usepackage{graphicx}
\usepackage{times}
\usepackage{url}
\newcommand{\Teff}{$\rm T_{eff}$}
\begin{document}

\Pagespan{1}{}
\Yearpublication{2014}%
\Yearsubmission{2013}%
\Month{1}%
\Volume{335}%
\Issue{1}%
\DOI{This.is/not.aDOI}%

\title{High resolution study of the abundance pattern of the heavy elements 
in very metal-poor field stars.
}

\author{M. Spite\inst{1}\fnmsep\thanks{Corresponding author.
  \email{monique.spite@obspm.fr}}
\and  F. Spite\inst{1}
}
\titlerunning{Abundance pattern of the heavy elements in very metal-poor field stars}
\authorrunning{M. Spite \& F. Spite}
\institute{
GEPI Observatoire de Paris, CNRS, Universit\'e Paris Diderot, F-92195
Meudon Cedex France}

\received{XXXX}
\accepted{XXXX}
\publonline{XXXX}


\abstract{
The abundances of heavy elements in EMP stars are not  well explained by the  simple view of an  initial basic 'rapid' process. In a careful and homogeneous analysis of the "First stars" sample (eighty per cent of the stars have a metallicity $\rm[Fe/H] \simeq -3.1\pm0.4$), it has been shown that at this metallicity [Eu/Ba] is constant, and therefore the europium-rich stars (generally called ``r-rich'') are also Ba-rich. 
The very large variation of [Ba/Fe] (existence of ``r-poor'' and ``r-rich'' stars) induces that the early matter was not perfectly mixed. 
On the other hand, the distribution of the values of [Sr/Ba]  vs. [Ba/Fe] appears with well defined upper and lower envelopes. No star was found with $\rm[Sr/Ba] < -0.5$ and the scatter of [Sr/Ba] increases regularly when [Ba/Fe] decreases.
To explain this behavior, we suggest that an early ``additional'' process forming mainly first peak elements would affect the initial composition of the matter. For a same quantity of accreted matter, this additional Sr production would barely affect the r-rich matter (which already contains an important quantity of Sr) but would change significantly the composition of the r-poor matter.
The abundances found in the CEMP-r+s stars reflect the transfer of heavy elements from a defunct AGB companion. But the abundances of the heavy elements in CEMP-no stars present the same characteristics as the the abundances in the EMP stars. 
Direct stellar ages may be found from radioactive elements, the precision is limited by the precision in the measurements of abundances from faint lines in faint stars, and the uncertainty in the initial abundances of the radioactive elements.  
}

\maketitle

\section{Introduction}

The detailed chemical composition of the old very metal-poor stars, formed in the early phases of the Galaxy, can provide unique clues to the complex formation of the heavy elements, and to the early history of the stellar  formation.

The heavy elements (here the elements from Strontium to Uranium) are built through neutron captures on seed nuclei (iron peak) by at least two different ways : the slow process (``s-process'') and the rapid process (``r-process''), which are slow or rapid relative to the beta decay time scale.
These processes are very different: in case of the ``s-process'' each seed receives about 1 neutron per 100 to 100\,000 years, but the ``r-process'' occurs when the seed receives more then 10 neutrons per second.

These two very different processes build the same elements (and often the same isotopes), but in very different ratios.

 The ``r-process'' requires an astrophysical environment with a so high neutron density that it cannot be reproduced in the laboratory yet. The astrophysical sites of the r-process are not yet clearly identified (Langanke \& Thielemann \cite {LT13}). It can occur only in cataclysmic events like the explosions following the core collapse of supernovae, or the merging of neutron stars or black holes, or the neutrino driven wind of a black hole torus... Its productions have been traced  very early in the Galactic evolution, and has enriched the matter forming the early metal-poor stars. 

The usual  ``s-process''  (called "main s-process") occurs frequently inside quite common stars, of the asymptotic giant branch, but only at the end of a long stellar evolution:  it becomes significant only at later time in the chemical evolution of the Galaxy. Since we are interested in old stars, formed early in the Galactic evolution, we have not to consider such a long duration process, but we have to be open to non-standard (or "weak") s-process"  occurring rapidly in massive fast evolving primitive stars (a possible late contamination in binaries by an evolved companion has also to be considered).

In stars formed more recently, like the Sun, the low mass AGB stars had time to enrich the interstellar (and the Sun-forming) matter by the elements of the "main "s-process" (observed in great detail in the Sun and meteorites).

\begin {figure}[h]
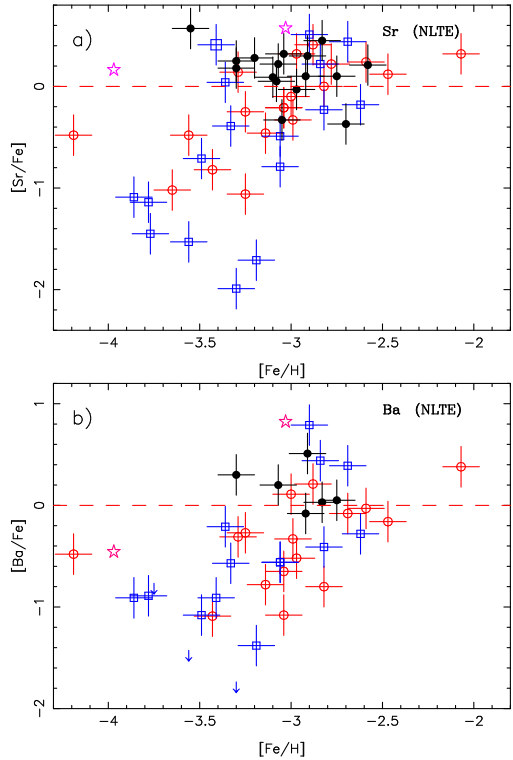

\begin {center}
\resizebox  {6.7cm}{5.0cm} 
{\includegraphics {abdgfe-srNLTE.ps} }
\resizebox  {6.7cm}{5.0cm} 
{\includegraphics {abdgfe-baNLTE.ps} }
\caption {
a) [Sr/Fe] and b) [Ba/Fe] vs. [Fe/H], for the sample of metal poor stars studied in the frame of the ESO large program ``First stars'' (Fran\c cois et al., \cite{FDH07}; Andrievsky et al., \cite{ASK11}). The open symbols represent giant stars (blue for the ``unmixed'' giants and red for the ``mixed'' giants: Spite et al. \cite{SCP05}, \cite{SCH06}) and the black filled circles represent turnoff stars. In both cases the scatter is very large. 
}
\label{obs1}
\end {center}
\end {figure}

\section{What is observed ?}
In the frame of the ESO large program ``First Stars'' we studied homogeneously a set of 50 very metal poor field stars: Cayrel et al. (\cite{CDS04}), Bonifacio et al. (\cite{BMS07}, \cite{BSC09}). Most of the stars have $\rm[Fe/H]<-3.0$ and are, therefore, ``extremely metal-poor'' (EMP) following  Beers \& Christlieb (\cite{BC05}). The S/N of the spectra is about 100 per pixel, with 5 pixels per resolution element, and with a  resolving power  $R\sim40\,000$.   

\subsection{Scatter of [Sr/Fe] and [Ba/Fe] vs. [Fe/H]}
\begin {figure}[h]
\begin {center}
\resizebox  {6.0cm}{4.5cm} 
{\includegraphics {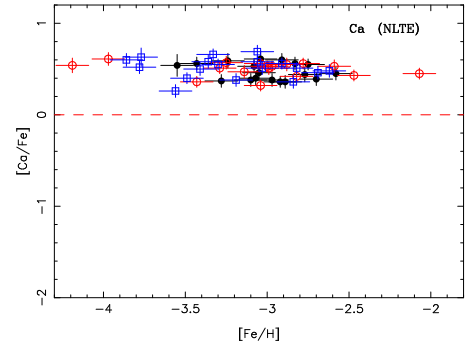} }
\caption {
 [Ca/Fe] vs. [Fe/H] at the same scale as Fig. \ref{obs1} for the same sample of stars with the same symbols.
}
\label{obs2}
\end {center}
\end {figure}

For this sample of stars, in Fig. \ref{obs1}a and \ref{obs1}b, [Sr/Fe] and [Ba/Fe] are plotted versus [Fe/H]. 
The scatter of [Sr/Fe] and [Ba/Fe] is huge compared to the scatter of, for example,  [Ca/Fe] vs. [Fe/H] (Fig. \ref{obs2}) plotted for comparison, at the same scale, for the same sample of stars.
In these figures the abundances of Sr, Ba and Ca have been corrected for NLTE effects (Andrievsky et al., \cite{ASK11}, Spite et al., \cite{SAS12}).

For a same metallicity [Fe/H], the ratios [Sr/Fe] or [Ba/Fe] can differ by a factor of almost 100.
A first question arises: is a barium-rich star also europium-rich, strontium-rich?

\begin {figure}[h]
\begin {center}
\resizebox  {5.0cm}{6.0cm} 
{\includegraphics {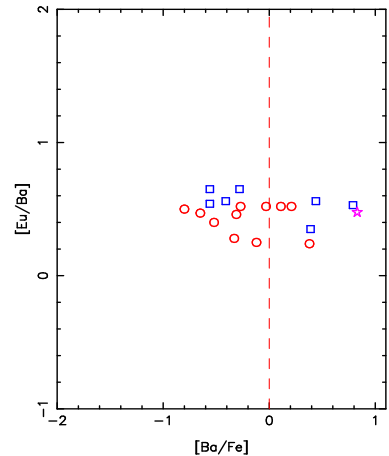} }
\caption {
 [Eu/Ba] vs. [Ba/Fe] for the same sample of stars as in Fig.\ref{obs1}. Europium can be measured in a limited number of EMP stars because the Eu lines, very weak, are not always measurable.
}
\label{obs3}
\end {center}
\end {figure}

\subsection{Behavior of [Eu/Ba] vs. [Ba/Fe]}  \label{euba}
If we plot for the same sample of stars, [Eu/Ba] as a function of the barium enrichment [Ba/Fe] (from Fran\c cois et al. \cite{FDH07}), the ratio [Eu/Ba] is constant and independent of the barium enrichment (Fig. \ref{obs3}). In these extremely metal-poor stars the europium lines are very weak and can be measured only in a limited number of stars.
Mashonkina et al. \cite{MCB10} found also that  [Ba/Eu] is constant in metal-poor stars as soon as $\rm[Fe/H]<-2.5$ (see their figure 12).
Since  [Eu/Ba] is constant in the most metal-poor stars, the Barium-rich stars are also Europium-rich.

Generally the ``r-rich'' stars (see Barklem et al. \cite{BCB05}) are defined relative to europium and are divided in two groups the r-II stars with  $\rm[Eu/Fe]_{LTE}>1$ and the r-I stars with  $\rm0.3<[Eu/Fe]_{LTE}<1$.  

The use of Europium  as a witness of the ``r-process'' is generally essential because Eu can be formed basically, only through  the ``r-process''. At intermediate metallicity ($\rm -1>[Fe/H]>-2.5$) in the Galactic evolution,  the low mass AGB stars had time to enrich the matter in Ba through the main ``s-process''. At these intermediate metallicities Eu, is a much better witness of the action of the ``r-process'' than Ba. But at  {\it very} low metallicity, even Ba has been formed by the ``r-process'' (unless the invocation of an early weak-s or non-standard s-process).    
 As a consequence, in EMP stars, a barium-rich star can be considered as a ``r-rich star''. To $\rm[Eu/Fe]_{LTE}>0.3$ corresponds $\rm[Ba/Fe]_{LTE}>-0.2$. 
All the Ba-rich stars of our sample where Eu can be measured are also Eu-rich, and accordingly belong to the class of the r-rich star: $\rm[Eu/Fe]_{LTE}>0.3$.

\begin {figure}[h]
\begin {center}
\resizebox  {6.0cm}{7.0cm} 
{\includegraphics {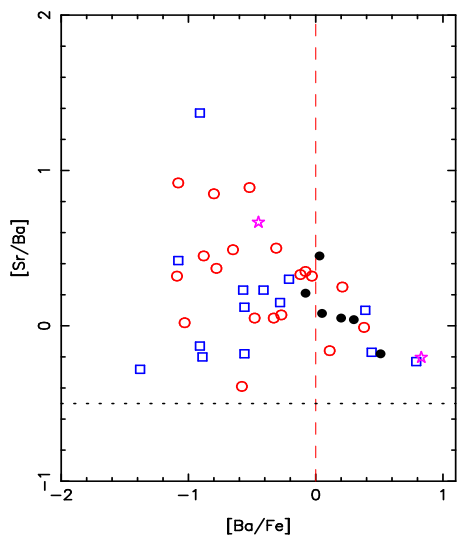} }
\caption {
 [Sr/Ba] vs. [Ba/Fe] for the same sample of stars as in Fig.\ref{obs1}.  Same symbols.
}
\label{obs4}
\end {center}
\end {figure}

\subsection{Behavior of [Sr/Ba] vs. [Ba/Fe]}
If now, we plot at the same scale, [Sr/Ba] as a function of [Ba/Fe] (Fig. \ref{obs4}), we see that the scatter of [Sr/Ba] is much larger for the Barium-poor stars than for the Barium-rich stars. Moreover this scatter increases regularly when [Ba/Fe] decreases and the slope of the upper envelope of the scatter is very close to one. 

If we understand the behavior of  [Eu/Ba] vs. [Ba/Fe] at low metallicity  (section \ref{euba}), it is more difficult to interpret the Fig. \ref{obs4} since Sr is expected to be produced in early events (massive stars or others) similarly to Eu and Ba. However, we remark that with Z=56 and Z=63,  Ba and Eu belong to the second peak of the r-elements while Sr with Z=38 belongs to the first peak. 
 It is then interesting to compare the detailed  abundance pattern of the heavy elements for different values of [Ba/Fe].


%

%

%


\section{Comparison of the abundance patterns of the r-elements in the EMP stars and in the Sun}

\subsection{Pattern of the heavy elements formed by the ``r-process'' in the Sun}
As already noted, the matter which formed the Sun has been enriched by the products of  massive and  
low-mass stars. As a consequence the heavy elements observed in the solar matter (solar atmosphere and meteorites) have been built by the ``r'' and the ``main s'' processes. 
It is however possible to compute the quantity of each element formed by the ``s'' process, because in the Sun:\\
-- the isotopic abundances are known (meteorites),\\
-- some isotopes, like $\rm ^{86}Sr, ^{134}Ba, ^{136}Ba, ^{154}Gd,$ can be built only through the ``s'' process,\\
-- the ``s-process'', which formed the heavy s-elements observed in the Sun, has occurred during the helium burning phase of AGB stars, and thus the physical conditions of the formation are known.\\

\begin {figure}[ht]
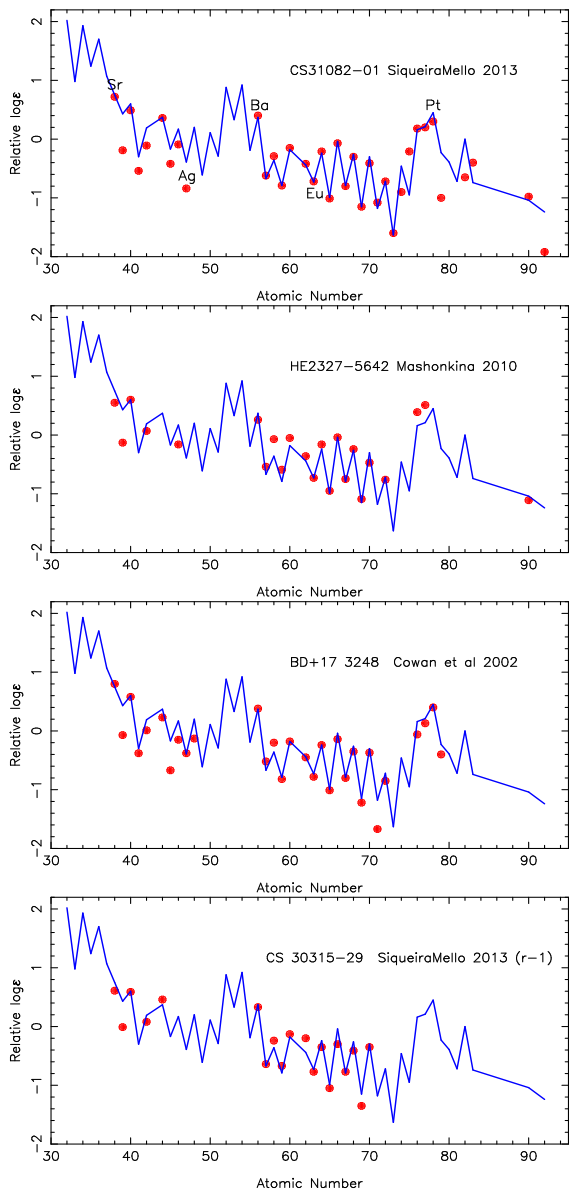

\begin {center}
\resizebox  {7.5cm}{3.9cm} 
{\includegraphics {cs31082-01.ps} }
\resizebox  {7.5cm}{3.9cm} 
{\includegraphics {he2327-5642.ps} }
\resizebox  {7.5cm}{3.9cm} 
{\includegraphics {bd17-3248.ps} }
\resizebox  {7.5cm}{3.9cm} 
{\includegraphics {cs30315-29.ps} }
\caption {
Abundance pattern of the heavy elements in four r-rich stars analyzed by different authors (red dots) are compared to the solar abundance pattern of the ``r'' elements following Simmerer et al. (\cite{SSC04}). CS31082-001 (Siqueira Mello et al., \cite{SSB13}), ~BD+17~3248 (Cowan et al. \cite{CSB02}), and ~HE2327-5642 (Mashonkina et al., \cite{MCB10}) are classified r-II and CS30315-29 r-I (Siqueira Mello et al., {\it in preparation}). Between Ba and Pt the abundance patterns of the r-rich stars are very similar to the solar abundance pattern.
}
\label{pat-rich}
\end {center}
\end {figure}

Then, by subtraction, it is possible to compute the quantity of each element built by the ``r-process''.
These computations have been done in particular, by Arlandini et al. (\cite{AKW99}), 
 Simmerer et al. (\cite{SSC04}), Bisterzo et al. (\cite{BGS11}), and the resulting solar r-process abundance patterns are very similar. 

\begin {figure}[ht]
\begin {center}
\resizebox  {6.0cm}{7.0cm} 
{\includegraphics {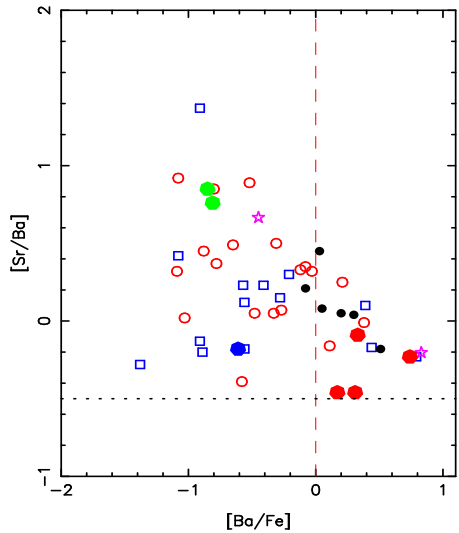} }
\caption {
 [Sr/Ba] vs. [Ba/Fe]  as in Fig.\ref{obs4}. The position of the r-rich stars of Fig. \ref{pat-rich} are marked with red filled circles and the position of the r-poor stars of Fig. \ref{pat-poor1} and \ref{pat-poor2} with green or blue filled circles. The lower envelope shows a [Sr/Ba] value compatible with the solar r-process ratio.
}
\label{srba-patt}
\end {center}
\end {figure}

\subsection{Comparison of the r-process abundance pattern in the Sun and in the r-rich stars}

The abundance pattern of the heavy elements in metal-poor stars has been first established for r-rich stars because it is more easily determined in r-rich stars which display stronger spectral lines.
In Fig. \ref{pat-rich} the abundance pattern of four r-rich stars are compared to the solar r-process pattern (Simmerer et al., \cite{SSC04}).
The position of these stars in the diagram [Sr/Ba] vs. [Ba/Fe] is shown in Fig. \ref{srba-patt} (filled red circles).
Between Ba and Pt (Fig. \ref{pat-rich}) the abundance patterns of the r-rich stars are closely similar to the solar abundance pattern. For this reason  it had been believed that the r-process was universal.
However, for the lightest elements, from Sr to Ag the agreement is not so good.

\begin {figure}[h]
\begin {center}
\resizebox  {7.5cm}{3.9cm} 
{\includegraphics {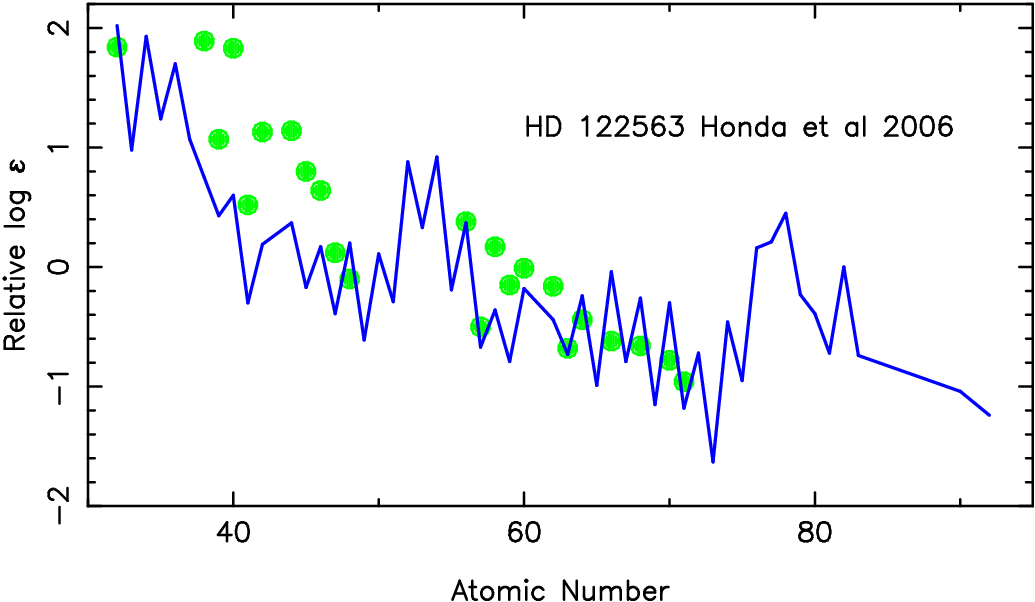} }
\resizebox  {7.5cm}{3.9cm} 
{\includegraphics {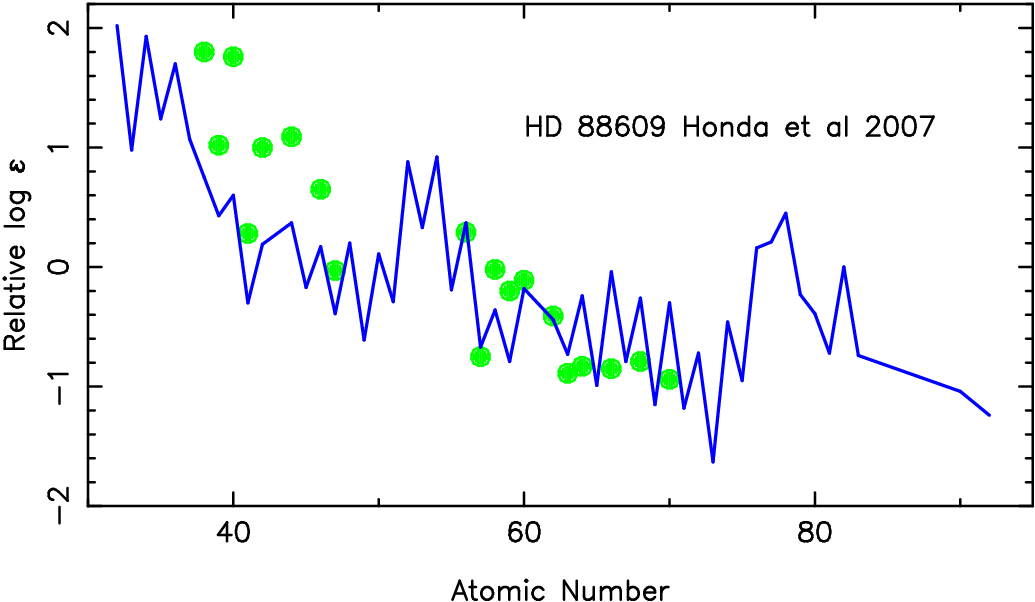} }
\caption {
Abundance pattern of the heavy elements in two r-poor stars: HD122563, (Honda et al., \cite{HAI06}, Roederer et al \cite{RLS12}) and HD88609 (Honda et al., \cite{HAI07}) compared to the solar abundance pattern following Simmerer et al. (\cite{SSC04}). 
In Fig. \ref{srba-patt} these stars with a high value of the ratio [Sr/Ba] are represented by green filled circles. 
}
\label{pat-poor1}
\end {center}
\end {figure}  

\subsection{Comparison of the r-process abundance pattern in the Sun and in the r-poor stars} 

The abundance pattern of the heavy elements for the r-poor stars is more difficult to establish and it could be defined on a reasonable set of lines only in two bright metal-poor stars  HD\,122563 (Honda et al., \cite{HAI06}; Roederer et al., \cite{RLS12}) and HD\,88609 (Honda et al., \cite{HAI07}), but on a smaller set in  BD--18:5550 (Fran\c cois et al., \cite{FDH07}).

HD~122563 and HD~88609 are characterized  by a high value of the ratio [Sr/Ba] (see Fig. \ref{srba-patt}, where these stars are represented by green filled circles), and on the contrary BD--18:5550 has a low [Sr/Ba] ratio (blue filled circle in Fig. \ref{srba-patt}). 

\subsubsection{r-poor stars with a high value of [Sr/Ba]}
The abundance pattern of the  heavy elements in HD~122563 and HD~88609 (Fig. \ref{pat-poor1}) differs significantly from the solar abundance pattern. The lightest heavy elements are strongly enhanced and even the second peak between Ba and Lu looks different.

\subsubsection{r-poor stars with a low value of [Sr/Ba]}
\begin {figure}[h]
\begin {center}
\resizebox  {7.5cm}{3.9cm} 
{\includegraphics {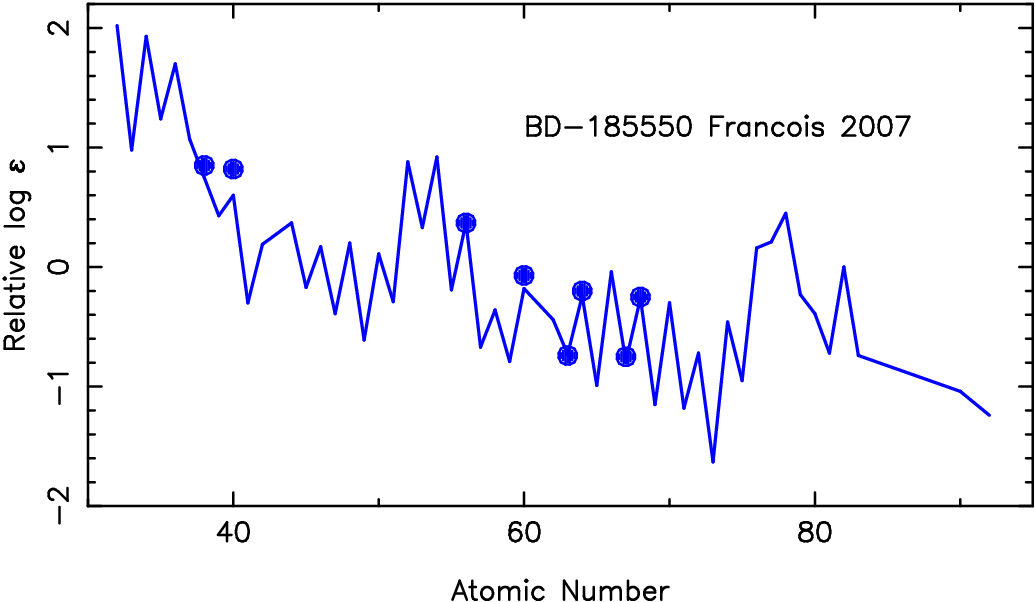} }
\caption {
Abundance pattern of the heavy elements in BD-18:5550 following Fran\c cois et al. \cite{FDH07} (red dots) compared to the solar abundance pattern following Simmerer et al. (\cite{SSC04}). In Fig. \ref{srba-patt} this star which has a low value of the ratio [Sr/Ba] is represented by a blue filled circle.
}
\label{pat-poor2}
\end {center}
\end {figure}

On the other hand, many Ba-poor stars like BD-18:5550 have a low value of the ratio  [Sr/Ba] like the Ba-rich stars and it would be interesting to compare the pattern of their heavy elements to the solar abundance pattern (Fig. \ref{pat-poor2}). In BD-18:5550, Fran\c cois et al. (\cite{FDH07}) could measure the abundance of eight heavy elements, and even if the abundance pattern of the heavy elements in this star is far to be complete, this pattern is rather similar to the solar abundance pattern  as it is for the Ba-rich star.

\subsubsection{A summary of the behavior of the heavy elements in EMP stars at [Fe/H]=--3}

Eighty per cent of the EMP stars studied homogeneously in the frame of the ESO large program ``First stars'' have $\rm -3.5<[Fe/H]<-2.7$, or $\rm[Fe/H] \simeq -3.1\pm0.4$ (see e.g. Fig. \ref{obs1}). They are thus representative of the behavior of the stars around [Fe/H]=--3.\\
--At this metallicity not a single star has been found with $\rm[Sr/Ba]<-0.5$, the solar r-process ratio,     (see also Aoki et al., \cite{ABL13}).\\
--The observations show that more the stars are r-rich and more the scatter of [Sr/Ba] is small: for [Ba/Fe]=+0.6, $\rm-0.5<[Sr/Ba]<-0.2$ (the scatter is about 0.3 dex, all the r-rich stars have a low [Sr/Ba] ratio), but for  [Ba/Fe]=--1.0, $\rm-0.5<[Sr/Ba]<+1.4$ (the scatter is about 1.9 dex).\\
--The pattern of the heavy elements abundance in EMP stars is not the same in all the EMP stars, and thus it cannot be considered as ``universal''.\\ 
--This pattern seems to be directly linked to the ratio [Sr/Ba] and does not directly depend 
on the fact that the star is r-poor or r-rich, i.e. does not depend on [Eu/Fe] or [Ba/Fe], but depends on the ratio [Sr/Ba]. Even the ratios of the second peak elements seem to be affected.\\

\section{Theoretical computations of the abundance patterns of the heavy elements in EMP stars}
  
There have been several attempts to fit the abundance pattern of the EMP stars with nucleosynthesis models. In spite of increasing observational data, the astrophysical site of this nucleosynthesis is still unclear.

For the {\bf r-process}, have been considered e. g.:\\
--nucleosynthesis during different phases of the explosion of core-collapse supernovae with a mass of about $20 M_{\odot}$: proto-neutron-star wind in core collapse SNe (Wanajo, \cite{Wan13}), cold r-process in neutrino driven wind (Wanajo, \cite{Wan07}; Farouqi et al., \cite{FKP10}; Thielemann et al., \cite{TAK11}).\\
--very massive core collapse supernovae with $M$ between $25 M_{\odot}$ and~ $40 M_{\odot}$ (Boyd et al., \cite{BFM12}) which lead to a truncated r-process which favors the formation of the first peak elements.\\
--nucleosynthesis in electron capture supernovae. These less massive supernovae $9 M_{\odot}$ explode after the carbon burning phase when the presence of  $\rm ^{22}Ne$ in the core induces the ejection of an important neutron flux (Wanajo, \cite{WJM11}). This process could be the source of additional formation of first peak elements.\\
--nucleosynthesis in neutrino driven wind after merging of neutron stars and formation of a black hole torus (Wanajo \& Janka, \cite{WJ12}). See also Goriely et al. (\cite{GBJ11}), and Korobkin et al. (\cite{KRA12}). 

Finally, heavy elements could be formed through a ``{\bf weak-s-process}'' or {\bf non-standard s-process} in fast rotating low metallicity ($\rm -7<[Fe/H]<-3$) massive stars ($M=25 M_{\odot}$).
Rotation induces a primary production of $\rm ^{22}Ne$ which leads to an important neutron flux through the reaction $\rm ^{22}Ne (\alpha, n) ^{25}Mg$. This could be an early source of the lightest heavy elements like Sr, Y, Zr, the production depending (significantly) on the mass, metallicity and rotation velocity of the star (see e.g. Frischknecht et al., \cite {FHT12}; Cescutti et al., \cite {CCH13}).\\ 


Following Barbuy et al. (\cite{BSH11}) and Siqueira Mello et al. (\cite{SSB13})    using extended data from spatial UV domain, the cold r-process in neutrino driven winds (Wanajo et al., \cite{Wan07}) leads to the best representation of the abundance pattern of the heavy elements in the r-rich stars (see in particular, the Fig. 15 of Siqueira Mello et al., \cite{SSB13}) although the computations overestimate the production of the heavy elements in a very narrow range of elements :  Nb and Ag (Z= 41 to 47), see also  Hansen et al. (\cite{HPH12}), Peterson (\cite{Pet13}).

The abundance pattern of a   r-poor star with a high value of the ratio [Sr/Ba] like HD122563 can be represented by a superposition of several entropy components with a moderately neutron-rich wind (Farouqi, \cite{Far09}) however the abundance of the heaviest elements of the second peak is strongly underestimated.\\
In fact a single theoretical process alone is not able to represent completely the abundance patterns of the heavy element with realistic values of the physical parameters.\\

Fig. \ref{pat-rich} and \ref{pat-poor2} suggest that a family of more or less similar processes (like the cold r-process in neutrino driven wind) would be responsible for the abundance pattern of the  heavy elements of the early metal-poor stars, with a pattern more or less similar to the solar pattern: r-rich stars but also r-poor stars with a low value of the Sr/Ba ratio.

Then another additional process would (sometimes) enrich this matter in heavy elements ejecting preferably, elements of the first peak, explaining the high Sr/Ba abundance in some Ba-poor stars, but ejecting also elements between Ba and Lu since the relative abundance of the elements in this region seems to be different in stars with a high and a low value of the ratio Sr/Ba.\\
Before such an enrichment, the ratio [Sr/Ba] is about -0.3  in the matter forming Ba-rich or Ba-poor stars. Addition of a given quantity of matter rich in first peak elements, would barely affect the abundance ratios of the matter responsible for the Ba-rich stars since it is also Sr-rich (compared to the matter forming Ba-poor stars). As a consequence the scatter of Sr/Ba in the r-rich star is small. On the contrary the matter forming the Barium poor stars would be strongly affected, their ratio [Sr/Ba] would increase depending on the quantity of this accreted new ejecta. This would explain that the scatter of [Sr/Ba] is larger in r-poor stars than in r-rich stars.

Some more work will be necessary for defining the production sites of the heavy elements. 
For example, it would be very important to confirm the abundance pattern of the heavy elements in Ba-poor stars with a low value of the ratio [Sr/Ba] to confirm that they have the same abundance pattern as the r-rich stars.

\begin {figure}[h]
\begin {center}
\resizebox  {7.5cm}{9.0cm} 
{\includegraphics {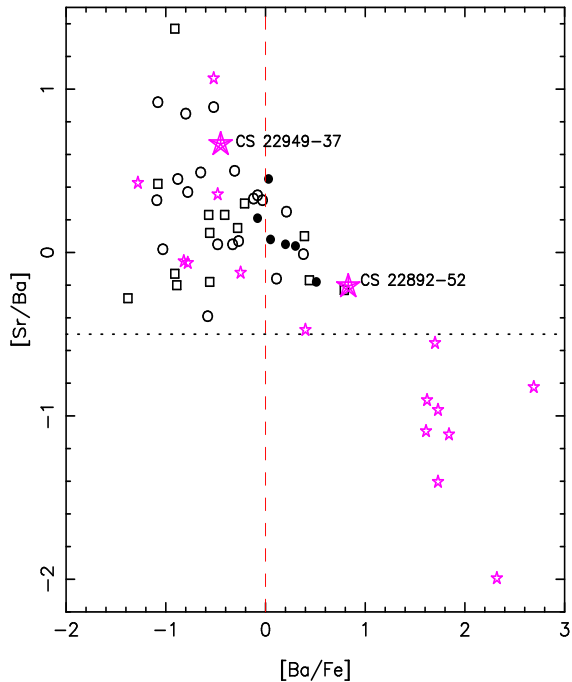} }
\caption {
 [Sr/Ba] vs. [Ba/Fe] for normal stars (black symbols, see Fig.\ref{obs4}) and carbon rich stars (pink stars symbols). The two big star symbols represent the two CEMP stars compared to similar EMP stars in Fig. \ref{com-pat-low} and Fig. \ref{com-pat-high}.
}
\label{obs4-C}
\end {center}{obs4-C}
\end {figure}

\begin {figure}[h]
\begin {center}
\resizebox  {7.5cm}{3.9cm} 
{\includegraphics {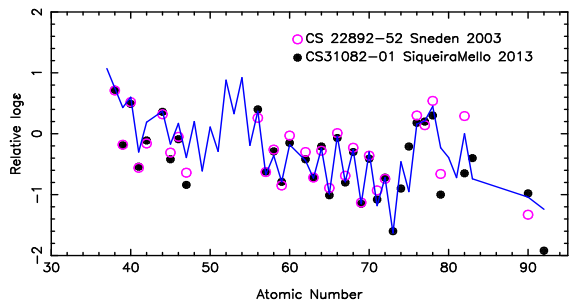} }
\caption {
Abundance pattern of the heavy elements in CS\,22892-052 (CEMP star, open pink circles) compared to the pattern of CS\,31082-001 (normal EMP star, black filled circles). Both stars have a low value of the ratio [Sr/Ba]. The solar abundance pattern following Simmerer et al. (\cite{SSC04}) is represented by a blue line. 
The abundance pattern of the two stars is very similar and thus does not depend on the carbon enrichment.
}
\label{com-pat-low}
\end {center}
\end {figure}

\section{The carbon-rich ``CEMP'' stars}
Many low metallicity ($\rm[Fe/H]<-4.0$) stars are carbon-rich i.e. have a ratio $\rm [C/Fe] > 1.0$ (called CEMP stars, following Beers \& Christlieb, \cite{BC05}). The number of CEMP stars increases when the metallicity decreases: they are  20\% at $\rm [Fe/H] \leq -2.0$ (Beers \& Christlieb, \cite{BC05}; Lucatello et al., \cite{LBC06}),  32\% at $\rm [Fe/H] \leq -3.0$ (Yong et al. \cite{YNB13b}), and below $\rm [Fe/H] = -4.5$, among the five stars which have been detected with this extremely low metallicity, four are C-rich; only SDSS~J102915+172927  (Caffau et al., \cite{CBF12}) is known to have a ``normal'' carbon abundance, i.e. $\rm [C/Fe]=+0.3\pm 0.3$ (Bonifacio et al., (\cite{BSC09}).\\
It is interesting to compare at very low metallicity the abundance pattern of the heavy elements in the normal metal-poor stars and in the carbon-rich stars.

In Fig. \ref{obs4-C} the position of the carbon-rich stars are compared to the normal stars in a diagram [Sr/Ba] vs [Ba/Fe]. The data are from Depagne et al. (\cite{DHS02}), Andrievsky et al. (\cite{ASK11}), Yong et al. (\cite{YNB13a}).\\ 
There are clearly two distinct groups of CEMP stars:\\ 

--stars with $\rm [Ba/Fe] > 1$ are all binaries and correspond to the CMP r+s stars (enriched in Ba and Eu). They are, as a mean, less metal-poor, they have a high value of the carbon abundance ($\rm A(C) \simeq 8.25$. The pattern of the heavy elements in these stars is very peculiar with a strong overabundance of Ba and Pb and the heavy elements are supposed to be mainly transferred from a defunct companion in its AGB phase although the process is not yet perfectly understood (Masseron et al., \cite{MJP10}).\\ 

\begin {figure}[h]
\begin {center}
\resizebox  {7.0cm}{3.75cm} 
{\includegraphics {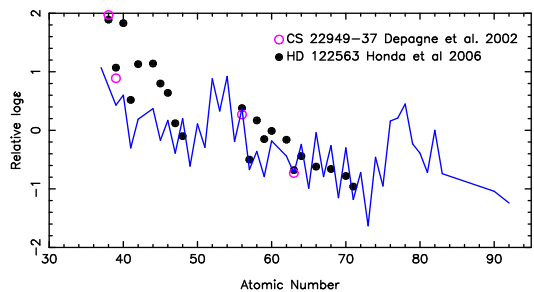} }
\caption {
Abundance pattern of the heavy elements in CS\,22949-037 (CEMP star, open pink circles) compared to the pattern of HD122563 (normal EMP star, black filled circles). Both stars have a high value of the ratio [Sr/Ba]. The solar abundance pattern following Simmerer et al. (\cite{SSC04}) is represented by a blue line. 
The abundance pattern of the two stars is compatible, but the abundance determination of more elements in CEMP stars with a high value of [Sr/Ba] is necessary to confirm this result.
}
\label{com-pat-high}
\end {center}
\end {figure}

--stars  with $\rm [Ba/Fe] < 1$ correspond to the CEMP-no stars (Masseron et al. \cite{MJP10}). They are not binaries and have a much lower value of the carbon abundance: $\rm A(C) \simeq 6.5$  (Spite et al., \cite{SCB13}). In Fig. \ref{obs4-C} these stars present the same wide distribution of [Sr/Ba] vs. [Ba/Fe] as the normal EMP stars, some have a low value of [Sr/Ba] and other a higher value.\\
-If we compare  a CEMP-no star CS\,22892-52 (Sneden et al., \cite{SCL03}), and a EMP star CS\,31082-001 (Siqueira Mello, \cite{SSB13}), with  {\it low} values of [Sr/Ba], it appears that the abundance patterns (Fig. \ref{com-pat-low}) are quite similar.\\  
-If we now compare (Fig. \ref{com-pat-high}) the abundance patterns of the heavy elements in EMP and CEMP-no stars with a {\it high} value of the ratio [Sr/Ba] : HD~122563 (Honda et al., \cite{HAI06})  and CS\,22949-037 (Depagne et al., \cite{DHS02}), here again the pattern of the heavy elements seem compatible. Unfortunately only few elements could be measured in CS\,22949-037, and the abundance determination of additional heavy elements in CS\,22949-37  (or any other CEMP star with a high value of [Sr/Ba]) would be useful to confirm the similarity of the patterns.

Finally,  no significant difference between the pattern of the heavy elements in the EMP and CEMP-no stars has been observed.

\section{Are there CEMP low-s stars at very low metallicity?}

In the samples of very metal-poor EMP and CEMP stars considered here, we have never observed stars with $\rm[Sr/Ba]< 0.5$ and $\rm[Ba/Fe]<+1$ (see Fig.\ref{obs4-C}). The low values of [Sr/Ba] are found only in carbon-rich stars ($\rm[C/Fe]>1.0$) and are associated with a very high value of [Ba/Fe]:  $\rm[Ba/Fe]>1$ (Fig.\ref{obs4-C}).\\
However Aoki et al. (\cite {ASB13}) have reported that they have found 5 stars with 
$\rm[Sr/Ba]<-0.5$. 
Three of these stars have  $\rm[Fe/H]<-2.8$ and can be thus directly compared to our star samples.\\
-- CS\,30322-023  (Masseron et al. \cite{MVF06}, Aoki et al. \cite{ABC13})\\
-- CS\,29493-090 (Barklem et al. \cite{BCB05}) \\
-- HE~0305-4520 (Barklem et al. \cite{BCB05})\\
The first of these stars has been studied in detail by Masseron et al. (\cite{MVF06}) and Aoki  et al. (\cite{ABC13}). On the other hand we had good high resolution spectra of the two other stars in the region 3350-8000\AA~ (Spite \& Spite in preparation), and we could measure the abundance of some clue elements.
These three stars are moderately Ba-rich ($\rm[Ba/Fe] \sim 0.5$) and seem to be also Pb-rich. But they are all Eu-poor: [Eu/Fe]=--0.63 for CS\,30322-023,  [Eu/Fe]=--0.23 for CS29493-90 and [Eu/Fe]=--0.30 for HE~0305-4520. 
All these stars are cool giants, they have the characteristics of mixed giants: the nitrogen abundance is very high and the ratio $\rm^{12}C/^{13}C$ is small. These stars could have been C-rich before mixing (their carbon has been little by little transformed into nitrogen).  
These three stars could belong to the class of the CEMP-low-s stars following Masseron et al. (\cite{MVF06}) and then the abundance pattern of the heavy elements would be the result of of mass transfer from a defunct AGB companion.\\
However, comparing the radial velocities of the two last stars, to the radial velocities of Barklem et al. (\cite{BCB05}) we did not find any indication of binarity. The stars were observed in 2001 by Barklem et al. (\cite{BCB05}), and in 2005 and 2006 for the new observations of CS\,29493-090 and HE~0305-4520. Another possibility would be that these stars be in fact evolved stars in their AGB phase.

\section{Is it possible to date a star ?}
Some of the heavy elements are radioactive and, therefore, if it is possible to measure the present-day abundance of this element $\rm A(R)_{now}$, if the half life of this element in Gyear $\rm \tau_{1/2}$ and its initial abundance   $\rm A(R)_{initial}$ are known, then it is possible to compute that R has been built $\rm \Delta t$ Gyear ago since 
$\rm A(R)_{now}= A(R)_{initial} ~e^{-Ln 2~ \Delta t / \tau_{1/2}} $. \\
If we suppose that the star has been formed immediately after the explosion of the supernova (immediate recycling), $\rm \Delta t$ is also the age of the star.
The initial abundance of a radioactive element is generally deduced from the comparison to the abundance of a stable element A(S): $\rm A(R)/A(S)_{now}= A(R)/A(S)_{initial} ~e^{-Ln 2~ \Delta t / \tau_{1/2}} $.   The ratio $\rm A(R)/A(S)_{initial}$ can be deduced from the solar r-elements pattern or a theoretical pattern. The age can be also deduced from the abundance ratio of two radioactive elements if the half lives of these elements are significantly different.\\

The abundance of two radioactive elements Th (Z=90) and U (Z=92) could be measured with a good precision in two EMP r-rich stars HE\,1523-0901 (Frebel et al., \cite{FCN07}) and CS\,31082-001 (Cayrel et al. \cite{CHB01}; Hill et al. \cite{HPC02}; Barbuy et al. \cite{BSH11}; Siqueira Mello et al. \cite{SSB13})\\

\noindent HE1523-0901\\
Using Eu as a reference stable element, Frebel et al. (\cite{FCN07}) deduced from [Th/Eu] an age between 9.5 and 13.3 Gyr depending on the adopted production ratio. From U/Eu they compute 13.2 Gyr and from U/Th an age between 12.2 and 13.9 Gyr. The final weighted average is $\rm 13.4 Gyr$.\\

\noindent CS31082-001\\
In this star the ratio Th/Eu is higher than in the Sun and thus the star appears to be younger than the Sun.
This can be explained only if this star as undergone a boost of the actinides (the star has been enriched in the third peak elements). 
If indeed the age of CS\,31082-001 is deduced from the ratio Th/Eu, both actinides, (see Barbuy et al., \cite{BSH11}) the age becomes $\rm 14.0 \pm 2.4$ Gyr.

Finally the ages of these two very old galactic stars are in perfect agreement with the age of the Universe deduced by the Planck collaboration (Ade et al., \cite{planck13}): $\rm 13.82 \pm 0.05$.






\begin{thebibliography}{}

\bibitem[2013]{planck13}
 Ade, P.A.R., Aghanim, N., Armitage-Caplan, C. et al. (Planck Collaboration), 2013, A\&A in press, 
arXiv:1303.5076

\bibitem[2011]{ASK11}
Andrievsky, S. M., Spite, F., Korotin, S. A., Fran\c cois, P., Spite, M., Bonifacio, P., Cayrel, R., Hill, V., 2011, A\&A 530, A105

\bibitem[2007]{ABC13}
Aoki, W., Beers, T. C., Christlieb, N., Norris, J. E., Ryan, S. G., Tsangarides, S., 2007, ApJ 655, 492

\bibitem[2013a]{ASB13}
Aoki, W., Suda, T., Boyd, R. N., Kajino, T., Famiano, M. A., 2013, ApJ 766, L13

\bibitem[2013b]{ABL13}
Aoki, W., Beers, T. C., Lee, Y. S., Honda, S., Ito, H., Takada-Hidai, M., Frebel, A., Suda, T., Fujimoto, M. Y., Carollo, D., Sivarani, T., 2013, AJ 145, 13

\bibitem[1999]{AKW99}
Arlandini, C., K\"appeler, F., Wisshak, K., Gallino, R., Lugaro, M., Busso, M., Straniero, O., 1999, ApJ 525, 886

\bibitem[2011]{BSH11}
Barbuy, B., Spite, M., Hill, V., Primas, F., Plez, B., Cayrel, R., Spite, F., Wanajo, S., Siqueira Mello, C., Andersen, J., 2011, A\&A 534, 60 ({\bf "First Stars XV''})

\bibitem[2005]{BCB05}
Barklem, P.S., Christlieb, N., Beers, T. C., Hill, V., Bessell, M.S., Holmberg, J., Marsteller, B., Rossi, S., Zickgraf, F.-J., Reimers, D., 2005, A\&A 439, 129

\bibitem[2005]{BC05}
Beers, T. C., Christlieb, N., 2005, ARAA 43, 531

\bibitem[2011]{BGS11}
Bisterzo, S., Gallino, R., Straniero, O., Cristallo, S., K\"appeler, F., 2011, MNRAS 418, 284

\bibitem[2007]{BMS07}
Bonifacio, P., Molaro P., Sivarani, T., Cayrel, R., Spite, M., Spite, F., 
Plez, B., Andersen, J., Barbuy, B., Beers, T. C., Depagne, E., Hill, V., 
Fran\c cois, P., Nordstr\" om, B., Primas, F., 2007, A\&A 462, 851
({\bf "First Stars VII''})

\bibitem[2009]{BSC09}
Bonifacio, P., Spite, M., Cayrel, R., Hill, V., Spite, F., Fran\c cois, P., Plez, B., 
Ludwig, H.-G., Caffau, E., Molaro, P., Depagne, E., Andersen, J., Barbuy, B., 
Beers, T. C., Nordstr\" om, B., Primas, F., 2009, A\&A  501, 519 ({\bf "First Stars XII''})

\bibitem[2012]{BFM12}
Boyd, R. N., Famiano, M. A., Meyer, B. S., Motizuki, Y., Kajino, T., Roederer, I. U., 2012 ApJL 744, L14

\bibitem[2000]{BPA00}
Burris, D. L., Pilachowski, C. A., Armandroff, T. E., Sneden, C., Cowan, J. J., Roe, H., 2000, ApJ 544, 302

\bibitem[2012]{CBF12}
Caffau, E., Bonifacio, P., Fran\c cois, P., Spite, M., Spite, F., Zaggia, S., Ludwig, H.-G., Steffen, M., Mashonkina, L., Monaco, L. et al., 2012, A\&A 542, A51

\bibitem[2001]{CHB01}
Cayrel, R., Hill, V., Beers, T. C., Barbuy, B., Spite, M., Spite, F., Plez, B., Andersen, J., Bonifacio, P., Franois, P., Molaro, P., Nordstr\"om, B., Primas F., 2001, Nature 409, 691 

\bibitem[2004]{CDS04}
Cayrel, R., Depagne, E., Spite, M., Hill, V., Spite, F., Francois, P., Plez, B., 
Beers, T., Primas, F., Andersen, J., Barbuy, B., Bonifacio, P., Molaro, P., 
Nordstr\"om, B., 2004, A\&A 416, 1117
({\bf "First Stars V''})

\bibitem[2013]{CCH13}
Cescutti, G., Chiappini, C., Hirschi, R., Meynet, G., Frischknecht, U., 2013, A\&A 553, A51

\bibitem[2002]{CSB02}
Cowan, J. J., Sneden, C., Burles, S., Ivans, I. I., Beers, T. C., Truran, J. W., Lawler, J. E., Primas, F., Fuller, G. M., Pfeiffer, B., Kratz, K.-L., 2002, ApJ 572, 861 
 
\bibitem[2002]{DHS02} 
Depagne, E., Hill, V., Spite, M., Spite, F., Plez, B., Beers, T. C., Barbuy, B., Cayrel, R., Andersen, J., Bonifacio, P., et al, 2002, A\&A 390, 187 ({\bf "First Stars II''})
 
\bibitem[2009]{Far09}
Farouqi, K., 2009, Proceedings of the 6th Russbach Workshop on Nuclear Astrophysics, 
\url{http://www.uni-mainz.de/Organisationen/vistars/russbach2009_lectures.html}  
 
\bibitem[2010]{FKP10}
Farouqi, K.; Kratz, K.-L.; Pfeiffer, B.; Rauscher, T.; Thielemann, F.-K.; Truran, J. W., 2010, ApJ 712, 1359
 
\bibitem[2007]{FDH07} 
Fran\c cois, P., Depagne, E., Hill, V., Spite, M., Spite, F., Plez, B., Beers, T. C.,
 Andersen, J., James, G., Barbuy, B., Cayrel, R., Bonifacio, P., Molaro,
P., Nordstr\"om, B.,Primas, F., 2007, A\&A 476, 935   ({\bf First stars VIII})

\bibitem[2007]{FCN07} 
Frebel, A., Christlieb, N., Norris, J.E., Thom, C., Beers, T.C., Rhee, J., 2007, ApJ 660, L117

\bibitem[2012]{FHT12} 
Frischknecht, U.,  Hirschi, R.,  Thielemann, F.-K., 2012,  A\&A 538, L2

\bibitem[2011]{GBJ11} 
Goriely, S., Bauswein, A., Janka, H.-T., 2011, ApJ 738, L32

\bibitem[2012]{HPH12} 
Hansen, C. J., Primas, F., Hartman, H., Kratz, K.-L., Wanajo, S., Leibundgut, B., Farouqi, K., Hallmann, O., Christlieb, N., Nilsson, H., 2012, A\&A 545, A31

\bibitem[2002]{HPC02} 
Hill, V., Plez, B., Cayrel, R., Beers, T. C., Nordstr\"om, B., Andersen, J., Spite, M., Spite, F., Barbuy, B., Bonifacio, P., Depagne, E., Fran\c cois, P., Primas, F., 2002, A\&A 387, 560  ({\bf First stars I})

\bibitem[2006]{HAI06} 
Honda, S.,  Aoki, W., Ishimaru, Y., Wanajo, S., Ryan, S.G., 2006, ApJ 643, 1180

\bibitem[2007]{HAI07} 
Honda, S.,  Aoki, W., Ishimaru, Y., Wanajo, S., 2007, ApJ 666, 1189

\bibitem[2012]{KRA12} 
Korobkin, O., Rosswog, S., Arcones, A., Winteler, C., 2012, MNRAS 426, 1940

\bibitem[2013]{LT13} 
Langanke, K.,  Thielemann F.-K., 2013, ENews 44/3, 24 

\bibitem[2013]{LBC06} 
Lucatello, S., Beers, T. C., Christlieb, N., Barklem, P. S., Rossi, S., Marsteller, B., Sivarani, T., Lee, Y. S., 2006, ApJ 652, L37  

\bibitem[2010]{MCB10}
Mashonkina, L.,  Christlieb, N., Barklem, P. S., Hill, V., Beers,T. C.,  Velichko, A., 2010, A\&A 516, A46

\bibitem[2006]{MVF06}
Masseron, T., van Eck, S., Famaey, B., Goriely, S., Plez, B., Siess, L., Beers, T.C., Primas, F., Jorissen, A., 2006, A\&A, 455, 1059

\bibitem[2010]{MJP10}
Masseron, T., Johnson, J. A., Plez, B., van Eck, S., Primas, F., Goriely, S., Jorissen, A., 2010, A\&A 509, A93 

\bibitem[2013]{Pet13}
Peterson, R. C., 2013, ApJ 768, L13

\bibitem[2008]{PGM08} 
Pignatari, M., Gallino, R., Meynet, G., Hirschi, R., Herwig, F., Wiescher, M., 2008, ApJ 687, L95

\bibitem[2012]{RLS12}
Roederer, I. U., Lawler, J. E., Sobeck, J. S., Beers, T. C., Cowan, J. J., Frebel, A., Ivans, I. I., Schatz, H., Sneden, C., Thompson, I. B., 2012, ApJS 203, 27

\bibitem[2004]{SSC04} 
Simmerer, J., Sneden C., Cowan, J. J., Collier J., Woolf V. M., Lawler J. E., 2004, ApJ 617, 1091

\bibitem[2013]{SSB13}
Siqueira Mello Jr., C., Spite M., Barbuy, B., Spite, F., Caffau, E., Hill, V., Wanajo, S., Primas, F., Plez, B., Cayrel, R., Andersen, J., Nordstr\"om, B., Sneden, C., Beers, T. C., Bonifacio, P., Fran\c cois, P., Molaro, P., 2013, A\&A 550, A122 ({\bf "First Stars XVI''})

\bibitem[2003]{SCL03}
Sneden, C., Cowan, J, J., Lawler, J, E., Ivans, I, I., Burles, S,, Beers, T, C., Primas, F, Hill, V, Truran, J, W., Fuller, G, M., et al., 2003, ApJ 591, 936

\bibitem[2005]{SCP05}
Spite, M., Cayrel, R., Plez, B., Hill, V., Spite, F., Depagne, E., Fran\c cois, P., Bonifacio, P., Barbuy, B., Beers, T., Andersen, J., Molaro, P., Nordstr\"om, B., Primas, F., 2005, A\&A 430, 655 ({\bf "First Stars VI''})

\bibitem[2006]{SCH06}
Spite, M., Cayrel, R., Hill, V., Spite, F., Fran\c cois, P., Plez, B., Bonifacio, P., Molaro, P.,  Depagne, E., Andersen, J.,  Barbuy, B.,  Beers, T. C.,  Nordstr\"om, B.,  Primas, F., 2006, A\&A 455,261 ({\bf "First Stars IX''})
 
\bibitem[2012]{SAS12}
Spite, M., Andrievsky, S. M., Spite, F., Caffau, E., Korotin, S. A., Bonifacio, P., Ludwig, H.-G., Franois, P., Cayrel, R.,  2012, A\&A 541, A143

\bibitem[2013]{SCB13}
Spite M., Caffau E., Bonifacio P., Spite F., Ludwig H.-G., Plez B., and Christlieb N., 2013, A\&A 552, A107

\bibitem[2011]{TAK11}
Thielemann, F.-K., Arcones, A., K\"appeli, R., Liebend\"orfer, M., Rauscher, T., Winteler, C., Fršhlich, C., Dillmann, I., Fischer, T., Martinez-Pinedo, G., Langanke, K., Farouqi, K., Kratz, K.-L. et al., 2011, PrPNP 66, 346   

\bibitem[2007]{Wan07}
Wanajo, S., 2007, ApJ 666, L77

\bibitem[2013]{Wan13}
Wanajo, S., 2013, ApJ 770, L22

\bibitem[2011]{WJM11}
Wanajo, S., Janka, H.-T., M\"uller, B., 2011, ApJ 726, L15

\bibitem[2012]{WJ12}
Wanajo, S., Janka, H.-T., 2012, ApJ 746, 780

\bibitem[2013b]{YNB13b}
Yong, D., Norris, J. E., Bessell, M. S., Christlieb, N.; Asplund, M.; Beers, Timothy C.; Barklem, P. S.; Frebel, Anna; Ryan, S. G., 2013, ApJ 762, 27

\bibitem[2013a]{YNB13a}
Yong, D., Norris, J. E., Bessell, M. S., Christlieb, N.; Asplund, M.; Beers, Timothy C.; Barklem, P. S.; Frebel, Anna; Ryan, S. G., 2013, ApJ 762, 26

\end{thebibliography}
\end{document}